\documentclass[10pt,conference]{IEEEtran}

\usepackage[table]{xcolor}
\usepackage[most]{tcolorbox}
\usepackage{url}

\IEEEoverridecommandlockouts

\usepackage{cite}
\usepackage{balance}
\usepackage{amsmath,amssymb,amsfonts}
\usepackage{algorithmic}
\usepackage{graphicx}
\usepackage{textcomp}
\usepackage{xcolor}
\usepackage{multirow}

\def\BibTeX{{\rm B\kern-.05em{\sc i\kern-.025em b}\kern-.08em
    T\kern-.1667em\lower.7ex\hbox{E}\kern-.125emX}}

\makeatletter
\newcommand{\linebreakand}{%
  \end{@IEEEauthorhalign}
  \hfill\mbox{}\par
  \mbox{}\hfill\begin{@IEEEauthorhalign}
}
\makeatother

\begin{document}

\title{LAURA: Enhancing Code Review Generation with Context-Enriched Retrieval-Augmented LLM}


\author{\IEEEauthorblockN{Yuxin Zhang}
\IEEEauthorblockA{\textit{School of Computer Science and Technology} \\
\textit{Beijing Institute of Technology}\\
Beijing, China \\
yuxinzhang@bit.edu.cn}
\and
\IEEEauthorblockN{Yuxia Zhang\IEEEauthorrefmark{2}}
\IEEEauthorblockA{\textit{School of Computer Science and Technology} \\
\textit{Beijing Institute of Technology}\\
Beijing, China \\
yuxiazh@bit.edu.cn}
\and
\IEEEauthorblockN{Zeyu Sun}
\IEEEauthorblockA{\textit{Institute of Software} \\
\textit{Chinese Academy of Sciences}\\
Beijing, China \\
zeyu.zys@gmail.com}
\linebreakand
\IEEEauthorblockN{Yanjie Jiang}
\IEEEauthorblockA{\textit{School of Computer Science} \\
\textit{Tianjin University}\\
Tianjin, China \\
2990094974@qq.com}
\and
\IEEEauthorblockN{Hui Liu}
\IEEEauthorblockA{\textit{School of Computer Science and Technology} \\
\textit{Beijing Institute of Technology}\\
Beijing, China \\
liuhui08@bit.edu.cn}

\thanks{\IEEEauthorrefmark{2}Corresponding author.}
}

\maketitle

\begin{abstract}
Code review is critical for ensuring software quality and maintainability. With the rapid growth in software scale and complexity, code review has become a bottleneck in the development process because of its time-consuming and knowledge-intensive nature and the shortage of experienced developers willing to review code. Several approaches have been proposed for automatically generating code reviews based on retrieval, neural machine translation, pre-trained models, or large language models (LLMs). These approaches mainly leverage historical code changes and review comments. However, a large amount of crucial information for code review, such as the context of code changes and prior review knowledge, has been overlooked. This paper proposes an LLM-based review knowledge-augmented, context-aware framework for code review generation, named LAURA. The framework integrates review exemplar retrieval, context augmentation, and systematic guidance to enhance the performance of ChatGPT-4o and DeepSeek v3 in generating code review comments. Besides, given the extensive low-quality reviews in existing datasets, we also constructed a high-quality dataset. Experimental results show that for both models, LAURA generates review comments that are either completely correct or at least helpful to developers in 42.2\% and 40.4\% of cases, respectively, significantly outperforming SOTA baselines. Furthermore, our ablation studies demonstrate that all components of LAURA contribute positively to improving comment quality.
\end{abstract}


 \section{Introduction}
\label{section:introduction}

Code review, a process of having peers manually examine source code changes, is critical to software development. Both open source and industrial software projects conduct code review activities to identify defects and ensure software maintainability \cite{bavota2015four, mcintosh2014impact, sadowski2018modern}. With the increasing growth of software scale and complexity, the drawbacks of modern code reviews, e.g., being time-consuming, cannot be overlooked, and many developers face a substantial review workload \cite{zhou2017scalability}. For example, in 2016, a core developer of the Linux kernel raised the argument \cite{JonathanCorbet}: ``\textit{We are getting more reviewers, but they are coming in slowly and are not anywhere near enough. As a result, the number of unprocessed patches is on the increase.}''  To alleviate this burden, a substantial amount of studies has focused on automating the code review process, such as recommending the best reviewers \cite{thongtanunam2015should, sulun2019suggesting, asthana2019whodo, chueshev2020expanding, mirsaeedi2020mitigating, chouchen2021whoreview, kong2022recommending, pandya2022corms}, predicting code locations that need review \cite{hellendoorn2021towards}, suggesting relevant code review comments \cite{hong2022commentfinder, shuvo2023recommending}, and optimizing code before submission \cite{tufano2021towards, tufano2022using}.

Driven by the rapid development of artificial intelligence technologies, many studies have shifted towards automatically generating code reviews by leveraging neural machine translation and pre-trained models \cite{tufano2022using, li2022automating, lu2023llama}. However, achieving satisfactory results remains challenging because of the complexity and knowledge-intensive nature of code review. 
The rise of large language models (LLMs) brings new opportunities for improvement, as they demonstrate exceptional performance in natural language understanding, generation, and a wide range of SE automation tasks \cite{eliseeva2023commit, sun2024gptscan, guo2024exploring}.  
However, directly using LLMs for code review generation faces several limitations. On one hand, LLMs often lack access to essential contextual information such as pull request (PR) details, code change specifics, and 
the domain-specific review experience that human reviewers rely on -- all of which are crucial for producing high-quality reviews \cite{kononenko2016code, turzo2024makes}. 
On the other hand, LLMs lack systematic guidance when generating code reviews; they typically cannot understand well the review’s scope and focus, nor do they operate within a well-defined logical framework to structure their feedback. This absence of context and structured guidance can lead to suboptimal review quality. 

To overcome these limitations, we propose \textbf{LAURA}, a novel \underline{\textbf{L}}LM \underline{\textbf{AU}}gmentation Framework for Code \underline{\textbf{R}}eview Gener\underline{\textbf{A}}tion. LAURA consists of three components -- context augmentation, review exemplar retrieval, and systematic guidance -- combining augmentation and prompts to improve LLMs in code review generation.
We leverage relevant contextual information of PRs and code changes 
and construct a history code review dataset 
to provide LLMs with relevant background knowledge and review experience. 
Specifically, since existing code review datasets do not have the context information we require and tend to have a quarter of low-quality data \cite{tufano2024code}, we construct a new dataset with careful filtering.
Our dataset contains 301,256 diff-comment-info series, which includes complete PR data from 1,807 popular GitHub projects. We manually annotate 384 recent entries for evaluation, while the remaining 298,494 entries dated before December 26, 2024, are used as the retrieval-augmented generation (RAG) database to provide valuable reference reviews. This enables the LLM to leverage relevant context and review experience, similar to a human reviewer. To address the second limitation -- lack of guidance -- we design prompts that provide clear instructions and a logical review structure, helping LLMs better understand review focus and generate coherent, high-quality comments.

We apply the LAURA framework to two well-performed LLMs, i.e., ChatGPT-4o and DeepSeek v3 (hereinafter, we name them as LAURA-GPT and LAURA-DS, respectively), and conduct experiments to validate the effectiveness of LAURA. 
To overcome the limitations of syntactic similarity-based metrics in evaluating LLM-generated text \cite{fiske1992but}, we conduct both LLM-based and human evaluations. The results show that LAURA-GPT and LAURA-DS consistently rank at the top across four out of five LLM evaluation metrics for assessing the overall review generation performance -- readability, relevance, sufficiency, and operability -- only slightly trailing in brevity. Additionally, both models achieve I-Score and IH-Score values of 20.0\%, 42.2\% and 18.5\%, 40.4\% respectively, in human evaluation metrics, which measure the usefulness of the generated content, meaning that both models effectively identified issues in about one-fifth of cases and provided at least helpful review comments in over two-fifths of cases. Compared to their respective base models, LAURA-GPT and LAURA-DS improve their I-Score by 30.5\% and 36.6\%, and their IH-Score by 37.3\% and 38.4\%, respectively. This indicates a significant improvement over directly using ChatGPT-4o and DeepSeek v3. LAURA also significantly outperforms CodeReviewer \cite{li2022automating}, the pre-trained SOTA approach in code review generation. Furthermore, we conduct an ablation study to analyze the impact of LAURA's three components on code review quality, confirming the effectiveness of the approach.

In summary, this paper makes the following contributions:
\begin{itemize}
\item We design the first composite method, which aims to improve the performance of LLMs in automatic code review generation by context augmentation, review exemplar retrieval, and systematic guidance.
\item We construct a high-quality dataset through hybrid filtering methods.
\item Our proposed method outperforms the state of the art in automatic code review generation. 
\item We have made the retrieval-augmented data publicly available to facilitate future improvements in code review generation \cite{LAURAfigshare}. 
\end{itemize}

\section{Approach Design}
\label{section:methodology}
\begin{figure*}[tbp]
  \centering
  \includegraphics[width=0.95\linewidth]{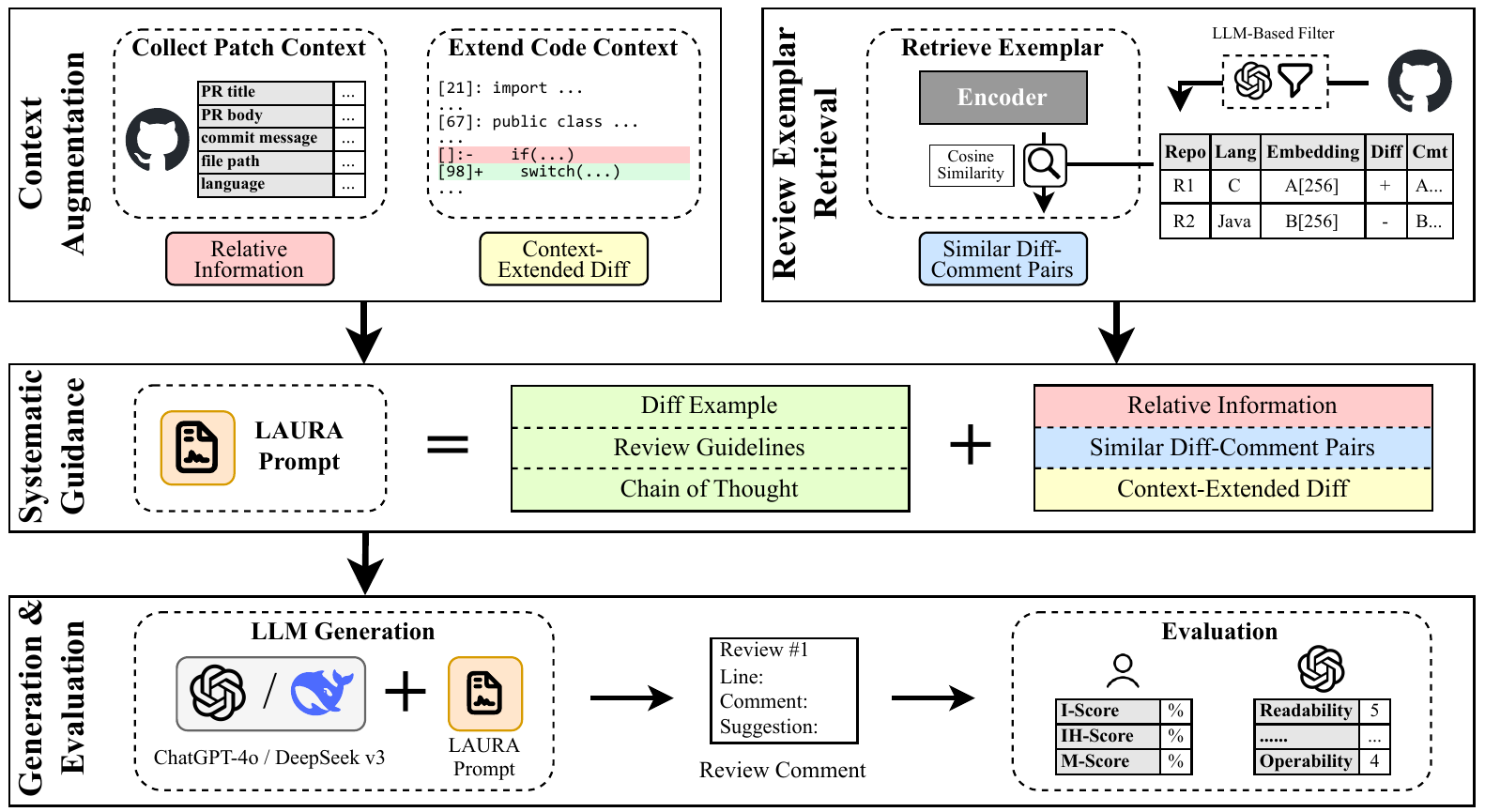}
  \caption{Overview of the LAURA framework.}
  \label{fig:methodOverview}
\end{figure*}

In this study, we propose an enhanced LLM approach to generate review comments for a given code change. This section provides a detailed description of the three components we designed to enhance LLMs, i.e., context augmentation, review exemplar retrieval, and systematic guidance. Figure~\ref{fig:methodOverview} shows an overview of our approach.

\subsection{Context Augmentation}
\label{subsection:informationAugmentation}

In a pull request (PR), the core is code changes, which are the object for code review. However, other information in a pull request can also play an important role in understanding the rationale of these code changes. 
In practice, a reviewer can read the PR title and body to understand the primary goal of the code change. 
The reviewer can then use the commit message and file path to identify the general purpose and content of the change, and finally refer to the code context to fully understand the modification 
in the diff. All of this contextual information played an important role in helping the reviewer thoroughly understand the code changes and provide meaningful review feedback. We believe that providing this information can help LLMs generate more helpful comments. 


However, existing studies only utilize code or diffs for code review generation. 
Meanwhile, since diffs by default only include three lines of context before and after, code context is often missing or insufficient. Therefore, in the context augmentation component, we integrate information that was mainly ignored before but is helpful to understand code changes and evaluate code issues. Specifically, we consider using the following information as a supplement and enhancement to code changes.

\textbf{Pull request title.} The PR title is a concise summary of the changes and often the first thing reviewers read to grasp the basic idea. Providing it to LLMs may help them infer the intent behind the changes and enhance their understanding.

\textbf{Pull request body}. The PR body typically explains the purpose, scope, and rationale of the changes, providing a context for the review. Supplying these details to LLMs can improve their overall understanding of the changes and help them more accurately identify potential issues.

\textbf{Commit message}. Each commit within a PR may cover different aspects of the changes, with messages documenting those changes \cite{tian2022what}. Commit messages typically contain less information but are more fine-grained. Including the commit message can provide LLMs with more commit-specific contextual information and help them better understand the changes.

\textbf{Changed file path}. Similarly, the file path provides more fine-grained information compared to the commit message, hinting at the file’s role and the nature of the changes. Prior work shows paths can aid in review tasks \cite{shuvo2023recommending}. It could help LLMs assess change types and locate issues more effectively.

\textbf{Code Context}. Closely tied to the diff, code context is crucial for discovering issues that can only be identified by considering the surrounding code, which is usually the most crucial information for identifying issues within the diff. Since diffs include limited code context by default, extending the code context is very important.

For each code change, we gather the relevant information mentioned earlier and provide all of it, except for code context, directly to the LLMs. We integrate the code context into the diff as follows: (1) We use Tree-sitter \cite{Tree-sitter} to build an Abstract Syntax Tree (AST) of the source file corresponding to the diff. If possible, we extend the diff to the boundaries of the enclosing function/method, limited to three times the original diff length; otherwise, we extend it up to that length. (2) We extract all header/library/package imports appearing before the diff and include any not already in the expanded diff. (3) We annotate each line of the extended diff with line numbers: added lines as ``[line number]+'', unchanged lines as ``[line number] '', and deleted lines as ``[]-''. In rare cases (under 3\%) where the source file retrieved via the GitHub GraphQL API doesn’t match the diff exactly, we only apply step (3). This approach enables us to create context-extended diffs, giving LLMs richer and more useful code context information.

Ultimately, we provide the LLMs with both the non-code context information and the enriched diff, enabling them to better review the code changes and generate more reliable review comments by leveraging these enhanced details.

\subsection{Review Exemplar Retrieval}
\label{subsection:reviewExampleRetrieval}

Existing studies indicate that the types of issues identified during code reviews and the level of usefulness of the reviews are related to the experience and knowledge of reviewers \cite{kononenko2016code, turzo2024makes}. Simply relying on LLMs can hardly grasp the pertinent knowledge from the vast amount of pre-training data. Thus, we hypothesize that providing LLMs with diffs similar to the code diff under review, along with the review comments made by other reviewers for those diffs, can serve as a reference for generating review comments. Prior studies \cite{wang2021context, zhang2024rag} applied retrieval-augmented approaches to automatically generate commit messages and achieved promising effectiveness. In this study, we first introduce the retrieval-augmented generation (RAG) method to retrieve diffs similar to the code diff under review and their corresponding review comments, using this retrieved information as part of the input to LLMs to achieve better quality in the generated review comments. 

We use cosine similarity scoring \cite{ochiai1957zoogeographic}, based on transformer encoders and vector embeddings, to assess code diff similarity and retrieve the most relevant code diffs and their review comments. For embedding, we use CodeT5+ \cite{wang2023codet5plus}, an enhanced version of CodeT5 \cite{wang2021codet5}, known for its Transformer encoder-decoder architecture and rich code representations from pre-training. We specifically use the ``codet5p-110m-embedding'' model to convert code diffs into 256-dimensional dense vectors, which encapsulate semantic information. Kartal et al. \cite{kartal2024automating} show that transformers, when pre-trained and fine-tuned, outperform traditional methods like Word2Vec and Code2Vec in encoding embeddings, making them well-suited for retrieval-augmented generation (RAG). After embedding the code diffs, we compute cosine similarity scores to identify the most relevant code diff and its review comments, which will be used to enhance the performance of LLMs.

Given that the diff to be reviewed may be extremely large, we introduce a token count threshold \textit{t} to balance cost and performance. We provide review exemplars for the diff under review as follows: (1) For diffs with token count no more than \textit{t}, we retrieve the most similar diffs (based on highest cosine similarity) from two sources -- exemplars from the same repository and exemplars in the same programming language. If the two retrieved diffs are not the same, we provide both sets of diff-comment pairs; otherwise, we only provide the pair from the same programming language. (2) For diffs with token count more than \textit{t}, we provide only the pair from the same programming language. We empirically set \textit{t} = 2,048 in our experiments. This approach offers more references for small diffs while avoiding overly long input contexts for large diffs.

\begin{figure*}[tbp]
  \centering
  \includegraphics[width=\linewidth]{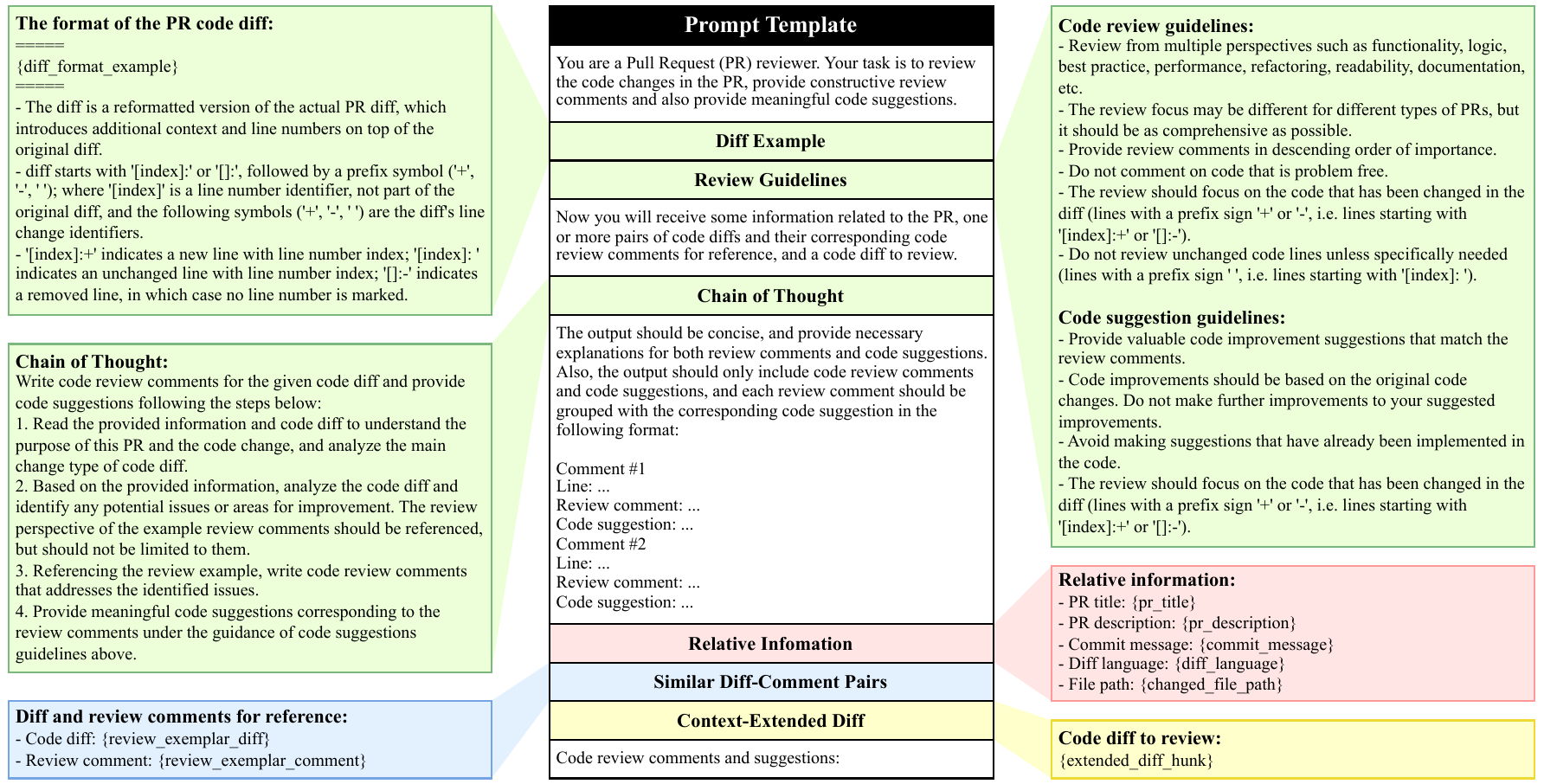}
  \caption{Prompts used for LAURA and direct generation.}
  \label{fig:prompts}
\end{figure*}

\subsection{Systematic Guidance}
\label{subsection:systematicGuidance}

We provide the augmented information, i.e., PR context and retrieved review exemplars, to LLMs along with the code diff under review through prompts. However, 
we still need to address the challenge that LLMs lack scientific guidance. Research shows that a well-designed prompt can guide LLMs in understanding the task and focusing on key points, leading to more structured, detailed, and accurate outputs, 
with its importance potentially surpassing that of the data itself \cite{white2023prompt}. Many automated tasks in software engineering have proved this, such as 
\cite{bae2024enhancing, ahmed2024automatic, pornprasit2024fine}. Thus, we also explore systematic guidance to enhance the performance of LLMs in the code review generation task. Inspired by relevant studies \cite{white2023prompt, amatriain2024prompt}, we design three prompt components -- 
diff example, review guidelines, and chain of thought -- to improve the generation quality of LLMs, focusing on input comprehension, task-critical point emphasis, and structured logical guidance, respectively.

\textbf{Diff example.} We provide an example of the extended code diff, along with a brief explanation of its components and format. Since our extension modifies the original diff structure by introducing longer code context and line numbers, we include this example and explanation to help LLMs better understand our input format and minimize potential misunderstandings.

\textbf{Review guidelines.} We design two sets of guidelines: review guidelines and code suggestion guidelines, to provide clearer instructions for LLMs in review generation. The review guidelines define the review’s scope and focus, while the code suggestion guidelines ensure LLMs provide compliant, useful suggestions, as human reviewers often do. We instruct LLMs to assess the code diff from multiple perspectives, emphasize key aspects based on the change type, and avoid redundant or meaningless suggestions. These guidelines aim to guide LLMs towards more effective reviews and prevent unproductive output.

\textbf{Chain of thought}. LLMs also need guidance on reasoning and information use. Prior work shows the chain of thought (CoT) approach can improve review outcomes \cite{tufano2024code}. Thus, we adopt CoT to provide LLMs with a clear logical framework, guiding them to logically process the augmented information: first, understanding the intent and content of changes, then identifying and explaining issues, and finally offering suggestions. We also integrate contextual information into the CoT to help LLMs better leverage relevant details during review.

Additionally, we provide LLMs with requirements for the generation format, as having a consistent generation format helps ensure that the outputs of LLMs are more organized and easier to read. The Prompt we used in our experiments is shown in Figure~\ref{fig:prompts}. LAURA uses the full prompt, while the direct generation method uses the main prompt with all additional parts removed.

\section{Experimental Design}
\label{section:experimentalDesign}

In this section, we present our experimental design, outlining the research questions we focus on, the process of constructing the dataset, the evaluation metrics and methods used, and the implementation details.

\subsection{Research Questions}
\label{subsection:researchQuestions}

To evaluate the impact of our proposed method on the performance of automatic code review generation, we design experiments to address the following two research questions.

\textbf{RQ1: How does LAURA perform on code review generation compared to the baselines?}
In this RQ, we evaluate our method using ChatGPT-4o and DeepSeek v3 to generate review comments on our evaluation dataset, as they represent closed-source and open-source LLMs, respectively. We compare two setups: (1) LAURA guidance, and (2) direct generation, and benchmark against CodeReviewer \cite{li2022automating}. Results are assessed via LLM and human evaluation.

\textbf{RQ2: How much does each of the three components of LAURA contribute to its overall effectiveness?}
In this RQ, we perform an ablation study, comparing three setups: (1) LAURA without context augmentation, (2) LAURA without review exemplar retrieval, and (3) LAURA without systematic guidance. Evaluation follows the same methods.

\subsection{Dataset Construction}
\label{subsection:datasetConstruction}

Only a few code review generation datasets are available, notably the T5CR dataset \cite{tufano2022using} and the CodeReviewer dataset \cite{li2022automating}. Many studies rely on the CodeReviewer dataset, but existing datasets suffer from quality issues. Tufano et al. found that around 25\% of these datasets are of low quality, with the CodeReviewer dataset \cite{tufano2024code} reaching about 32\%. This significantly limits dataset reuse. Furthermore, these datasets lack contextual information beyond the code itself and remove the source links of PRs, further constraining their usefulness. Therefore, we construct our own code review dataset with careful filtering to ensure high quality.

\subsubsection{Project Selection}

We selected high-quality open-source projects hosted on GitHub as case studies for our research. An overview of the project selection process is shown in Table~\ref{tab:dataInfo}. Specifically, we focused on projects primarily written in one of four programming languages: C, C++, Java, and Python. This selection covers a broad range of programming paradigms and includes languages that are widely used.
We first sorted these projects in descending order by the number of stars and retained only those with at least 2,500 stars. As a result, we obtained 6,467 repositories, with the number of repositories for C, C++, Java, and Python being 782, 1,089, 1,384, and 3,212, respectively. We then filtered for repositories with at least 500 pull requests, narrowing it down to 1,860, as those with fewer PRs may lack activity or proper code reviews. Finally, we conducted a manual review and removed the following types of projects: (1) tutorial-oriented projects; and (2) projects that were forked from other repositories. Ultimately, we selected 1,807 repositories, including 228 C, 420 C++, 346 Java, and 813 Python repositories. In the next step, we collected pull request information from these repositories.

\begin{table}[tbp]
  \caption{Statistics of dataset preparation.}
  \label{tab:dataInfo}
  \begin{center}
  \begin{tabular}{lr}
    \hline
    Procedure & Count \\
    \hline
    \rowcolor{gray!20}
    Project selection & \# Projects \\
    Preliminary selection of projects & 6,467 \\
    Filtered ( PR count \textless 500) & -4.607 \\
    Filtered (manual check) & -53 \\
    Remaining projects & 1,807 \\
    \hline
    \rowcolor{gray!20}
    Data collection and cleaning & \# Diff-comment-info series \\
    Preliminary collection of data & 1,020,537 \\
    Filtered (rule-based) & -202,704 \\
    Filtered (LLM-based) & -446,002 \\
    Merged (comments on the same diff) & -70,575 \\
    Remaining data & 301,256 \\
    \hline
  \end{tabular}
  \end{center}
\end{table}

\subsubsection{Data Collection}

Table~\ref{tab:dataInfo} summarizes our data collection and cleaning process. We primarily used the GitHub GraphQL API \cite{GitHubGraphQL} to retrieve pull request data, as its flexible query system allows precise JSON data retrieval via HTTP POST requests. We collected PR titles, bodies, numbers, states, authors, reviewers, review comments (with timestamps), commit messages, diffs, and file paths -- collectively referred to as a ``diff-comment-info series.'' 
We filtered out cases where the PR author and reviewer were the same. Since comment-associated diffs were truncated, we fetched complete diffs via the GitHub commit-comparing function \cite{GitHubCompare}.

To improve data quality, we applied the 10-line rule \cite{bosu2015characteristics}, which assumes that code changes within 10 lines of a review location likely address the review and thus indicate valuable feedback. Following Rong et al. \cite{rong2024distilling}, we sorted commits and reviews chronologically within PRs and checked whether later commits modified code near earlier review locations. We kept only data conforming to the 10-line rule. After initial processing, we obtained 1,020,537 diff-comment-info series, but further filtering was necessary as the 10-line rule alone does not guarantee high quality.

\subsubsection{Data Filtering}

After data collection, we cleaned the collected data. First, we applied rule-based filtering: (1) We removed bot-generated comments by filtering reviewer names with ``bot'', ``-cr'', or ``gpt'' suffixes, and by using Golzadeh et al.’s bot list \cite{golzadeh2021ground}. (2) We filtered invalid and irrelevant comments by (a) removing comments with two or fewer words, as they typically offer little value \cite{dyer2013boa, schall2024commitbench}, and (b) applying Tufano et al.’s heuristic algorithm based on manually designed patterns \cite{tufano2022using}. (3) We removed comments containing non-ASCII characters. After these steps, we retained 817,833 diff-comment-info series.

Next, we applied an LLM-based filter to improve review quality. As Tufano et al. \cite{tufano2024code} note, 25\% of existing review comments remain low-quality despite basic filtering. We used GPT-4o mini to evaluate each comment, its corresponding diff, and a task prompt:
``Determine whether the review comments provided are valuable for improving the code diff. If the review comments point out problems in the code diff or give useful improvement suggestions, and are easy to understand, answer `yes'; if the review comments are irrelevant to the content of the code diff (such as only requiring it to be done in a separate PR or commit), or are difficult to understand or not clearly described, answer `no'. The answer should be one of `yes' or `no', without other content.''

To assess the feasibility and effectiveness of this filtering method, we manually reviewed 240 random samples (60 per language), identifying 65 low-quality comments (27.1\%). 
We treated high-quality comments as positive samples, and the LLM filter achieved an accuracy of 0.914 and a recall of 0.606. This indicates that although some high-quality samples may be filtered out, the proportion of low-quality samples in the final constructed dataset is significantly reduced. While not ideal for precise quality evaluation, the method is effective for filtering, as retaining low-quality comments is more harmful than discarding a few valuable ones. After this step, we had 371,831 diff-comment-info series.

Finally, we merged all review comments for the same code diff within the same pull request to provide richer reference comments for retrieval and evaluation. This yielded 301,256 diff-comment-info series. After data collection and cleaning, we finalized the context-enhancement component for our hybrid method. 

\subsubsection{Retrieval System Construction}

We split the constructed dataset by time and selected 298,494 diff-comment-info samples prior to December 26, 2024, as the retrieval-augmented generation database. To improve retrieval efficiency, we pre-embedded the code diffs for similarity comparison using the ``codet5p-110m-embedding'' model \cite{wang2023codet5plus}, generating 256-dimensional vectors. For test diffs, we embedded them with the same model and calculated cosine similarity scores against the RAG database. The most similar diffs and their review comments were merged into the test diff-comment-info series as review references for the LLM. With data partitioning and post-processing complete, we finalized the review exemplar retrieval component for our composite method.

\subsubsection{Evaluation Dataset Construction}

Since LAURA involves LLM augmentation and human evaluation, large-scale experiments aren’t feasible. We intended to sample 384 instances for evaluation (95\% confidence level, 5\% margin of error). Given the possibility of residual low-quality data in the LLM-filtered set and the risk of bias from discarding valuable comments, we opted to manually annotate high-quality data from the dataset without LLM filtering. 
We manually annotated high-quality code review comments based on the following criteria: (1) Readability and comprehensibility (this criterion excluded comments like ``s/Check less/Check if less/'' that were difficult to understand); (2) Clear relevance to issues in the code under review (this criterion excluded comments with low relevance to the code itself, such as ``I don't think this should be part of this PR?''); (3) Containing at least an effective issue explanation or code suggestion (this criterion excluded comments with low informational value, such as ``When does this happen?'' or ``Thanks''); (4) Self-contained understanding (this criterion excluded comments like ``Ditto, also change this'' that lacked essential contextual information). The annotation continues until we obtain 384 valid samples. To minimize data leakage risk (i.e., the chance that LLMs had seen the data during training), we only selected the diff-comment-info series dated after December 26, 2024 (the release date of DeepSeek v3, the latest model used in our experiments). We merged and shuffled the data. 
Then, the first two authors independently annotated each sample. A Cohen's kappa \cite{cohen1960coefficient} of 0.883 between their results, indicating a high level of inter-rater agreement. Subsequently, any disagreements were resolved by conducting several meetings and involving a third author as arbiter. In total, we annotated 546 samples, yielding 384 high-quality samples for evaluation.

\subsection{Baselines}
\label{subsection:baseline}

Tufano et al. \cite{tufano2024code} compared ChatGPT-3 with T5CR \cite{tufano2022using}, CodeReviewer \cite{li2022automating}, and CommentFinder \cite{hong2022commentfinder} in code review generation/recommendation, concluding that SOTA methods outperform ChatGPT-3. Among these methods, \textbf{CodeReviewer} is also designed for reviewing code diffs. Therefore, we choose CodeReviewer as a baseline to evaluate our method against traditional pre-trained models. Additionally, since LAURA is based on \textbf{ChatGPT-4o} and \textbf{DeepSeek v3}\footnote{We exclude DeepSeek R1 due to insufficient data after its release date.}, we also select ChatGPT-4o and DeepSeek v3 as additional baselines to evaluate the performance improvements introduced by our method.

\subsection{Evaluation}
\label{subsection:evaluation}

Given the nature of LLMs and the diversity of natural language \cite{fiske1992but}, it is impractical to expect LLM outputs to closely match human reviewer comments. Thus, traditional metrics like BLEU and ROUGE are not suitable \cite{reiter2018structured}. To efficiently evaluate the overall performance of the review comments generated by LAURA, we adopt an LLM-based evaluation approach. However, LLM evaluation has limitations when it comes to assessing the practical usefulness of review comments in real-world scenarios. Therefore, we also introduce a human evaluation approach.

\subsubsection{LLM Evaluation}

We follow the LLM-based DeepCRCEval method by Lu et al. \cite{lu2025deepcrceval}, with some modifications: 
(1) merging some metrics to improve clarity; (2) instructing the LLM to rate each comment on a 1–5 scale according to five metrics (higher scores indicating better quality), instead of the original 10-point scale, since prior study \cite{joshi2015likert} shows that a 5-point scale is easier to understand and apply accurately, avoids the coarseness of a 3-point scale, and provides greater stability and consistency than a 10-point scale; (3) adding simple scoring guidelines for each metric. For example, for the ``Readability'' metric, we provided the LLM with the following explanation: ``In the 1-5 scoring system, a score of 5 means the text is completely clear and readable; a score of 3 means the expressed meaning can be basically understood; and a score of 1 means the text is nearly unreadable or incomprehensible.'' We believe these scoring guidelines help calibrate scores across different evaluation samples and positively impact the consistency of the LLM. Table~\ref{tab:LLMEvalMetric} shows the final metrics.

For RQ1, we provided ChatGPT-4o (specifically, GPT-4o-2024-11-20) with five anonymized sets of review comments per data sample -- generated by LAURA-GPT, LAURA-DS, ChatGPT-4o only, DeepSeek v3 only, and CodeReviewer -- and asked it to score them using the five metrics. For RQ2, due to LLM input length limits and the relative nature of our LLM-based approach, we split the evaluation. Each data sample was paired with two anonymized sets of five review comments: one from LAURA-GPT, its three ablations, and ChatGPT-4o only; the other from LAURA-DS, its three ablations, and DeepSeek v3 only. ChatGPT-4o scored each set using the same metrics.

\subsubsection{Human Evaluation}

Sometimes, LLM-based evaluation may not effectively assess the usefulness of the generated review comments in real-world scenarios. Therefore, we also conduct a human evaluation. 
Following Mahajan et al. \cite{mahajan2020recommending}, we adjust their method for LLM-generated results (Table~\ref{tab:manualEvalMetric}): we remove the ``Unavailable'' category, redefine the other three to suit our goals, and add an ``Uncertain'' category. This is because, in our evaluation, all test data have corresponding generated review comments, so the ``Unavailable'' case does not apply. 
However, given the limited knowledge of the repository and PR context, these review comments could be difficult to evaluate, leading to ``Uncertain'' cases. Specifically, since the data comes from 1,807 open-source projects, we cannot fully understand every project’s code requirements, organizational structure, or other contextual details. Therefore, even if we determine that the LLM-generated code review comments are generally useful, without factual grounding, we cannot be certain whether they would be accepted within the specific project context of the reviewed code. We categorize each generated review comment as ``Instrumental'', ``Helpful'', ``Uncertain'', or ``Misleading'', and use three metrics to evaluate quality.

\textbf{I-Score} ($ \frac{I}{I+H+U+M} \times 100\% $): Proportion of comments rated as ``Instrumental'', reflecting the ability to correctly identify and address issues in the code diff. Higher is better.

\textbf{IH-Score} ($ \frac{I+H}{I+H+U+M} \times 100\% $): Proportion of comments rated as ``Instrumental'' or ``Helpful'', indicating the overall usefulness of comments. Higher is better.

\textbf{M-Score} ($ \frac{M}{I+H+U+M} \times 100\% $): Proportion of ``Misleading'' comments that contain factual errors, incorrect suggestions, or fail to address the code diff. Lower is better.

\begin{table}[tbp]
  \caption{Adjusted LLM Evaluation Metrics.}
  \label{tab:LLMEvalMetric}
  \begin{center}
  \begin{tabular}{l|p{6.6cm}}
    \hline
    Metric & Description \\
    \hline
    Readability & Clear, easily understandable language. \\
    \hline
    Relevance & Directly related to the specific code. \\
    \hline
    Brevity & Clearly explain the issues identified, and at the same time, be concise and not lengthy. \\
    \hline
    Sufficiency & Cover all issues as much as possible, and point out the exact location of issues for comprehensive review. \\
    \hline
    Operability & Practical advice for addressing identified issues. \\
    \hline
  \end{tabular}
  \end{center}
\end{table}

\begin{table}[tbp]
  \caption{Adjusted Human Evaluation Categories.}
  \label{tab:manualEvalMetric}
  \begin{center}
  \begin{tabular}{l|p{6.0cm}}
    \hline
    Category & Description \\
    \hline
    Instrumental (I) & The review fully matches the ground truth, that is, identifying all issues or providing equivalent solutions, with no obvious errors. \\
    \hline
    Helpful (H) & The review partially matches the ground truth, that is, identifying some or all issues, or at least providing related suggestions, but may lack completeness or accuracy. \\
    \hline
    Uncertain (U) & The review does not overlap with the ground truth but offers suggestions on other aspects, with no errors or direct conflicts. Its usefulness is unclear based on the ground truth alone. \\
    \hline
    Misleading (M) & The review directly contradicts the ground truth, provides irrelevant or incomprehensible feedback, or offers no meaningful input. \\
    \hline
  \end{tabular}
  \end{center}
\end{table}


For RQ1, we recruited four volunteers, each with over five years of software development experience and programming proficiency. 
To manually assess the 384 results per method through a rigorous multi-stage annotation process: initial calibration with shared examples, a discussion to align scoring, and then independent ratings of the remaining dataset. First, each evaluator reviewed the initial 32 results per method together, learning from each other, gaining different perspectives, and aligning on scoring criteria. Then, the evaluators independently evaluated the next 96 results per method, achieving a Krippendorff’s Alpha of 0.804 with p \textless 0.001, indicating good agreement. Based on this consensus and considering evaluation efficiency, the remaining 256 diff-comment-info series were evenly distributed among the four evaluators. To minimize potential bias, we randomly shuffled the evaluation samples, which means the evaluators were blinded to comment sources. For RQ2, the same four evaluators evaluated the ablation experiment results on the 384 diff-comment-info series. Based on their evaluation experience in RQ1, the evaluators evenly divided the 384 evaluations, each handling 96 results. We report the ablation evaluation results for these 384 samples.

\subsection{Experimental Details}
\label{subsection:experimentalDetails}

The dataset was prepared on a platform with two NVIDIA GeForce RTX 3090 GPUs. During the data split step, we used the ``codet5p-110m-embedding'' model to embed code diffs, setting the ``max\_length'' to 2048 to balance input size and efficiency -- this setting ensured high efficiency while only 5.3\% of diffs were truncated. All other parameters remained at default values.

Experiments were conducted via the APIs of ChatGPT-4o (i.e., GPT-4o-2024-11-20) and DeepSeek v3, ensuring each data entry was tested independently with no external context beyond the designed prompts and inputs. We set the API temperature to 0.5, as this produced better outputs in preliminary tests. For CodeReviewer, another baseline model in RQ1, we performed additional fine-tuning using our custom dataset on the same platform. Since CodeReviewer had already been fine-tuned on its original dataset and demonstrated a strong understanding of the code review generation task, we further fine-tuned it on our dataset for 10 epochs, keeping all other parameters the same as those used in the original fine-tuning process.

\section{Experimental Results}
\label{section:experimentalResults}

In this section, we present the results of the effectiveness of LAURA and the impact of each of its three components. 

\begin{table*}[tbp]
  \caption{Evaluation results of code review generation.}
  \label{tab:RQ1Res}
  \begin{center}
  \begin{tabular}{l|ccccc|ccc}
    \hline
    \multirow{2}{*}{Method} & & \multicolumn{3}{c}{LLM Evaluation} & & \multicolumn{3}{c}{Human Evaluation} \\
    & Readability & Relevance & Brevity & Sufficiency & Operability & I-Score & IH-Score & M-Score \\
    \hline
    LAURA-GPT & \textbf{4.956} & \textbf{4.956} & \underline{4.185} & \textbf{4.859} & \textbf{4.953} & \textbf{20.0\%} & \textbf{42.2\%} & \textbf{0.8}\% \\
    ChatGPT-4o & 4.162 & 4.040 & \textbf{4.235} & 3.566 & 3.855 & 15.4\% & 30.7\% & 1.3\% \\
    LAURA-DS & \underline{4.761} & \underline{4.786} & 4.151 & \underline{4.556} & \underline{4.686} & \underline{18.5\%} & \underline{40.4\%} & \underline{1.0\%} \\
    DeepSeek v3 & 4.031 & 3.861 & 4.183 & 3.255 & 3.635 & 13.5\% & 29.2\% & \underline{1.0\%} \\
    CodeReviewer & 2.182 & 1.909 & 2.859 & 1.188 & 1.404 & 3.4\% & 10.2\% & 9.1\% \\
    \hline
    \multicolumn{9}{l}{The highest and second-highest results are in bold and underlined, respectively.}
  \end{tabular}
  \end{center}
\end{table*}

\subsection{RQ1: Effectiveness of LAURA}
\label{subsection:RQ1}

\begin{figure*}[tbp]
  \centering
  \includegraphics[width=\linewidth]{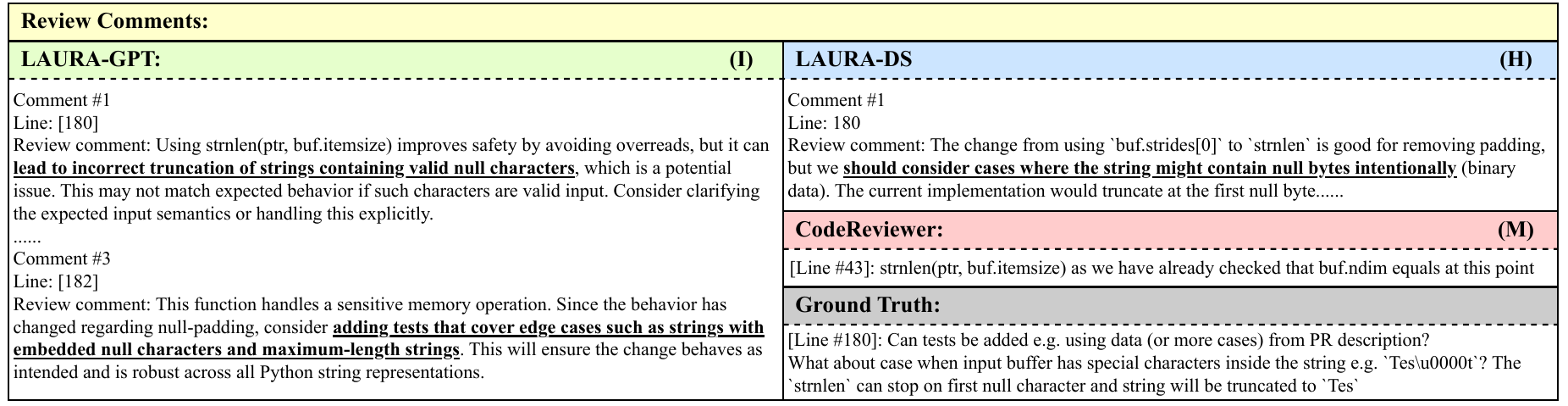}
  \caption{A real example of human evaluation of code review comments generated by LAURA-GPT, LAURA-DS, and CodeReviewer.}
  \label{fig:RQ1Example}
\end{figure*}

In this RQ, we ask ChatGPT-4o and DeepSeek v3 to review code diffs using the LAURA framework and also directly without it. Code diffs are also submitted to CodeReviewer for evaluation. Table~\ref{tab:RQ1Res} shows the results, reporting five LLM evaluation metrics, along with I-Score, IH-Score, and M-Score from human evaluation. The results show that LAURA-GPT achieves the highest I-Score of 20.0\% and IH-Score of 42.2\%, representing 30.5\% and 37.3\% improvements over directly using ChatGPT-4o, while also maintaining the lowest M-Score (0.8\%). It excels in four out of five LLM-evaluated metrics, with only a slight weakness in brevity. LAURA-DS ranks second, with an I-Score of 18.5\% and an IH-Score of 40.4\%, achieving improvements of 36.6\% and 38.4\% over directly using DeepSeek v3, respectively. Both methods perform well across most metrics, indicating that our method helps LLMs better identify areas for improvement in code changes. In contrast, CodeReviewer performs worst, with an I-Score of 3.4\%, IH-Score of 10.2\%, and the lowest scores in all LLM metrics, showing LLMs' superiority over traditional pre-trained models in realistic code review scenarios.

We provide a human evaluation example in Figure~\ref{fig:RQ1Example}. In this case, the diff under review modifies a string construction method, changing \textit{``std::string(ptr, buf.ndim == 0 ? buf.itemsize : buf.strides[0])''} to \textit{``std::string(ptr, strnlen(ptr, buf.itemsize))''}. The ground truth highlights a string truncation risk and suggests adding tests. LAURA-GPT identifies both issues, advises reverting the change, and recommends tests, earning an ``Instrumental'' rating. LAURA-DS detects the string truncation risk but omits the test suggestion, earning a ``Helpful'' rating. CodeReviewer simply approves the change without further suggestions, conflicting with the ground truth and receiving a ``Misleading'' rating.



\begin{figure*}[tbp]
  \centering
  \includegraphics[width=\linewidth]{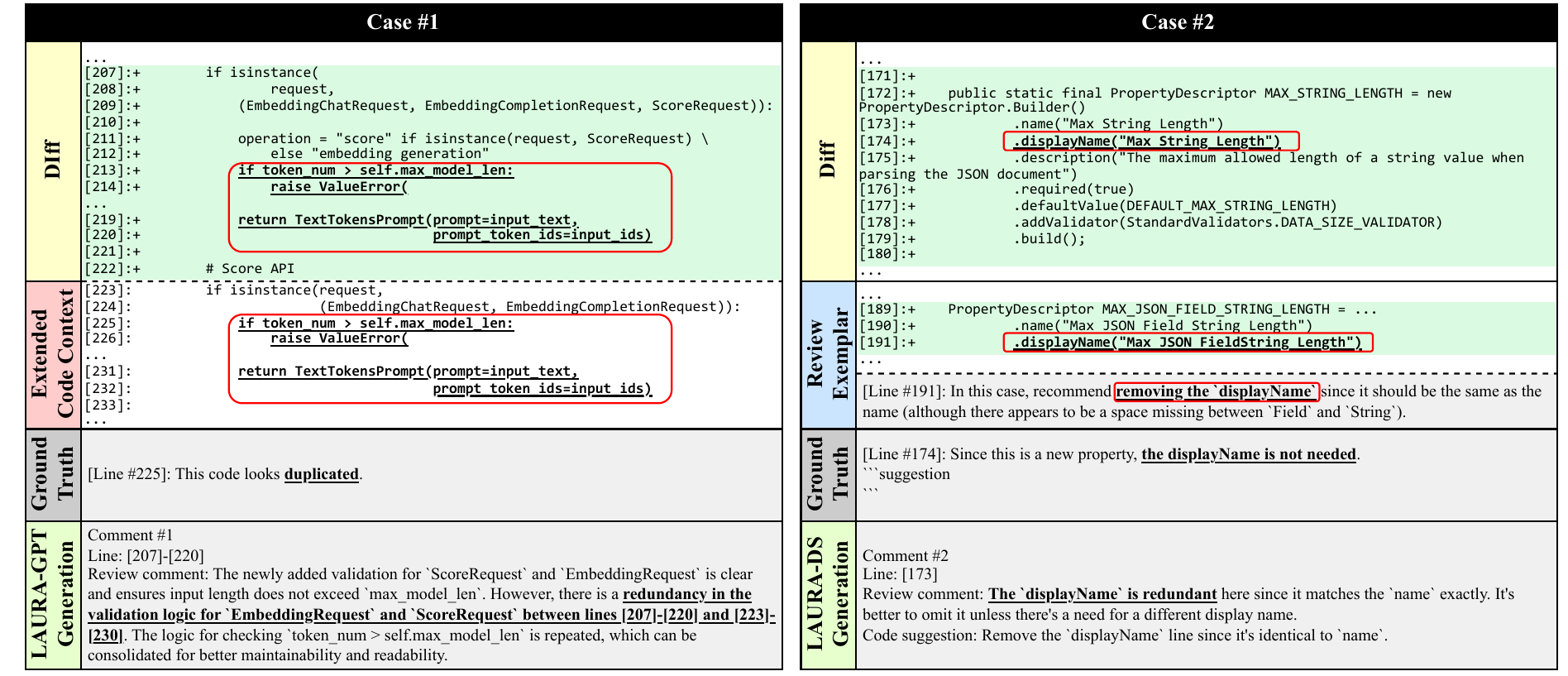}
  \caption{Two real cases of how LAURA assists LLMs in generating useful review comments.}
  \label{fig:RQ1Case}
\end{figure*}

The improvements are expected, as adding code change-related information equips LLMs with rich background knowledge, helping align their understanding with that of human reviewers. Retrieving similar code diffs and corresponding comments offers valuable reference experience, much like how human reviewers draw on past knowledge. 
As shown in Figure~\ref{fig:RQ1Case}, in case 1, the additional code context provided by LAURA highlights code duplication related to the diff, enabling GPT-4o to generate a review comment consistent with the ground truth. In case 2, the review exemplar provided by LAURA addresses the same issue of redundant ``displayName'' as the ground truth, thereby assisting DeepSeek v3 in generating a valuable review comment. These two real-world cases highlight the significant assistance provided by LAURA's helpful information in enabling LLMs to generate review comments.

In addition, in our experiments, the performance of directly using ChatGPT-4o for code review generation even surpasses that of CodeReviewer, which contrasts with the results of Tufano et al. \cite{tufano2024code}. We attribute this to advances in LLMs since ChatGPT-3, dataset differences, and varied evaluation methods. Our dataset, possibly containing more long code diffs, may have overwhelmed pretrained models. Additionally, Tufano et al. evaluated the three traditional methods as a whole, which also leads to differences in results.

\begin{center}
\begin{tcolorbox}[colback=gray!10,
                  colframe=black,
                  width=\linewidth,
                  arc=1mm, auto outer arc,
                  boxrule=0.3pt,
                  breakable,
                 ]
\textbf{RQ1 Summary:} LAURA significantly outperforms the traditional pre-trained model, CodeReviewer. Moreover, both LAURA-GPT and LAURA-DS achieve better performance compared to directly using the base model, indicating that this composite approach -- based on data augmentation and rapid fine-tuning -- effectively enhances the performance of LLMs in code review generation tasks.
\end{tcolorbox}
\end{center}

\subsection{RQ2: Ablation Study of the Three Key Components of LAURA}
\label{subsection:RQ2}

In this RQ, we instruct ChatGPT-4o and DeepSeek v3 to review code changes under conditions where one of the three components of LAURA is removed. Table~\ref{tab:RQ2Res} presents the experimental results, where we report the scores of comments generated by different methods on five metrics as evaluated by LLM, along with the I-Score, IH-Score, and M-Score of these comments as assessed by human evaluation. To make the comparison clearer, we also show the performance of LAURA-GPT and LAURA-DS with all three components in Table~\ref{tab:RQ2Res} shaded in light grey.

\begin{table*}[tbp]
  \caption{Evaluation results of the ablation study.}
  \label{tab:RQ2Res}
  \begin{center}
  \begin{tabular}{l|ccccc|ccc}
    \hline
    \multirow{2}{*}{Method} & & \multicolumn{3}{c}{LLM Evaluation} & & \multicolumn{3}{c}{Human Evaluation} \\
    & Readability & Relevance & Brevity & Sufficiency & Operability & I-Score & IH-Score & M-Score \\
    \hline
    \rowcolor{gray!20} LAURA-GPT & \textbf{4.956} & \textbf{4.956} & \textbf{4.185} & \textbf{4.859} & \textbf{4.953} & \textbf{20.0\%} & \textbf{42.2\%} & \textbf{0.8}\% \\
    w/o context augmentation & 4.393 & 4.341 & 4.140 & 3.838 & 4.101 & 17.7\% & 37.5\% & \textbf{0.8\%} \\
    w/o review exemplar retrieval & 4.561 & 4.522 & 4.166 & 4.163 & 4.314 & 18.2\% & 37.8\% & \textbf{0.8\%} \\
    w/o systematic guidance & 4.281 & 4.205 & 3.888 & 3.942 & 4.088 & 18.0\% & 37.2\% & \textbf{0.8\%} \\
    \hline
    \rowcolor{gray!20} LAURA-DS & \textbf{4.761} & \textbf{4.786} & \textbf{4.151} & \textbf{4.556} & \textbf{4.686} & \textbf{18.5\%} & \textbf{40.4\%} & \textbf{1.0\%} \\
    w/o context augmentation & 4.244 & 4.207 & 3.976 & 3.709 & 3.953 & 16.4\% & 35.9\% & \textbf{1.0\%} \\
    w/o review exemplar retrieval & 4.565 & 4.529 & 4.065 & 4.100 & 4.293 & 16.9\% & 37.0\% & \textbf{1.0\%} \\
    w/o systematic guidance & 4.321 & 4.309 & 3.929 & 3.916 & 4.110 & 16.4\% & 35.4\% & \textbf{1.0\%} \\
    \hline
    \multicolumn{9}{l}{The highest result in each group of results is highlighted in bold.}
  \end{tabular}
  \end{center}
\end{table*}

The ablation study results indicate that all three components of LAURA play important roles. Among them, the method that removes only the review exemplar retrieval component performs slightly better overall -- achieving the best LLM and human evaluation results within its respective ablation experiment group, whether based on ChatGPT-4o or DeepSeek v3. However, removing the review exemplar retrieval component also leads to a decrease in the I-Score of LAURA-GPT and LAURA-DS by 9.1\% and 8.4\%, respectively, and a decrease in the IH-Score by 10.5\% and 8.4\%, respectively. The methods that remove only the context augmentation component or only the systematic guidance component each have their own strengths and weaknesses, with no significant difference between them. However, all these methods show a decrease in LLM evaluation metrics compared to the complete method, while the I-Score in human evaluation dropped by 10.4\% - 11.7\%, and the IH-Score dropped by 11.0\% - 12.2\%.

These results indicate that all three components have a significant positive impact on improving the quality of LLM-generated outputs. This suggests that rich relevant information and highly related code context can indeed help LLMs better understand code changes, leading to more useful review comments. Providing exemplar reviews that reflect the experience of human reviewers can help LLMs more effectively assess review priorities and generate valuable feedback. Moreover, offering more detailed and precise instructions and guidance to LLMs can effectively enhance the usefulness of their comments.


Notably, LAURA’s performance gains in code review generation are not simply the sum of its three components, but the result of their interactions. The sum of the performance drop from removing each component individually exceeds the drop from removing all three (i.e., the baseline models). The slight advantage of removing only the review exemplar retrieval component, along with our evaluation observation that removing the systematic guidance component occasionally leads to comments targeting the exemplar instead of the diff, suggests that systematic guidance not only guides LLMs in reviewing but also helps them interpret the extended diff provided by the context augmentation component. Likewise, the extended diff helps LLMs better understand the review exemplars, enhancing the review exemplar retrieval component’s effectiveness. Together, these components reinforce each other, offering additional benefits to LLMs.

\begin{center}
\begin{tcolorbox}[colback=gray!10, 
                  colframe=black, 
                  width=\linewidth,
                  arc=1mm, auto outer arc,
                  boxrule=0.5pt,
                  breakable,
                 ]
\textbf{RQ2 Summary:} The three components of LAURA -- context augmentation, review exemplar retrieval, and systematic guidance -- all have a significant positive impact on the performance of LLMs. When used in combination, these components produce a compounding effect, further enhancing the relevance and usefulness of the review comments generated by LLMs.
\end{tcolorbox}
\end{center}

\section{Threats to Validity}
\label{section:threatsToValidity}

Threats to internal validity. The main threat to internal validity is the potential risk of data leakage. We sampled test data after the release of the latest model we used (DeepSeek v3) to avoid overlaps with the training data of LLMs. However, we cannot fully guarantee that the test data wasn’t used in model optimization post-release. Another threat is the length limit of extended diffs, as defining the optimal code context scope is challenging. Too little may miss details; too much may add noise, distract LLMs, and degrade output quality. Further research is needed to address this. In addition, the limitations of prompt exploration pose another threat to internal validity. Although we designed a composite prompt for generating code review comments based on prior research and experimental observations, there may exist better prompt designs and ordering strategies that could further enhance LAURA’s performance.

Threats to external validity. A threat to external validity lies in the quality of the dataset used during the research process. We used rule-based and LLM-based filters to remove low-quality comments, but some may remain. For the evaluation set, we relied on manual annotation to minimize this issue. Another threat to external validity concerns the choice of programming languages. Our experiments focused on C, C++, Java, and Python, so it’s unclear how well the results generalize to other languages. Our methods may serve as a reference for future research. 
Baseline selection may also pose a threat to external validity. Although we included open-source LLMs, closed-source LLMs, and state-of-the-art pretrained models as baselines, we did not incorporate some baselines due to factors such as differences in dataset structure and variations in programming language coverage. There is a promising avenue to evaluating LAURA on more LLMs.

Threats to construct validity. A threat to construct validity lies in the choice of LLMs. Although LAURA is model-agnostic, we used ChatGPT-4o and DeepSeek v3 as representative state-of-the-art models. This choice may influence results. Another threat to construct validity concerns the selection of evaluation metrics. LLM-based evaluation metrics have inherent variability across datasets, and while we mitigated this by merging metrics and adding scoring criteria, limitations persist. Our human evaluation focuses on basic factual alignment, so the practical usefulness of mismatched review comments remains uncertain.

\section{Related Work}
\label{section:backgroundAndRelatedWorks}

In this section, we introduce two areas related to our study: code review automation and large language models, and we discuss the relevant studies in each of these fields.

\subsection{Code Review Automation}
\label{subsection:codeReviewAutomation}


The code review process requires substantial time and effort from both parties. To improve efficiency, many studies have explored automating parts, including reviewer recommendation, review necessity prediction, identifying code areas for review, pre-review and post-review refinement, code review generation, and comment suggestion.

Early studies mainly focused on reviewer recommendation, using methods such as file path analysis \cite{thongtanunam2015should}, traceability graphs \cite{sulun2019suggesting}, load balancing \cite{asthana2019whodo}, collaborative filtering \cite{chueshev2020expanding}, and balancing expertise and workload \cite{mirsaeedi2020mitigating}. Other approaches considered cross-project collaboration \cite{kong2022recommending} and composite analyses of file paths, project details, and author info \cite{pandya2022corms}. Some research also aimed to predict which parts of the submitted code would need comments \cite{hellendoorn2021towards}.

Code review automation has advanced to include tasks like review necessity prediction, comment generation, and code refinement. Notable works include T5CR \cite{tufano2022using}, CodeReviewer \cite{li2022automating}, and CommentFinder \cite{hong2022commentfinder}. T5CR, based on the Text-To-Text Transfer Transformer (T5) model \cite{raffel2020exploring}, handles method-level code for pre-review optimization, comment generation, and post-review refinement. CodeReviewer uses T5 architecture with CodeT5 initialization \cite{wang2021codet5}, processes line-level diffs, and supports review necessity prediction, comment generation, and code refinement across nine languages. CommentFinder is an information retrieval (IR)-based method that uses Bag of Words (BoW) and similarity measures to find similar code snippets and recommend relevant review comments.



So far, several works have attempted to apply LLMs to code review automation tasks. Lu et al. \cite{lu2023llama} used the LLaMA model for review necessity prediction, comment generation, and code refinement, achieving results comparable to pre-trained models via parameter-efficient fine-tuning. Guo et al. \cite{guo2024exploring} showed that ChatGPT-3.5 outperformed CodeReviewer in post-review refinement. Pornprasit et al. \cite{pornprasit2024fine} explored ChatGPT-3.5’s performance in both pre- and post-review refinement, analyzing few-shot learning effects. Tufano et al. \cite{tufano2024code} compared ChatGPT-3 with three existing methods and found it achieved state-of-the-art performance in code refinement, but lagged in review generation. However, these studies provide only a limited exploration of LLMs and have not fully unlocked their potential for code review automation tasks.

In this work, we use line-level code diff hunks as the raw input and focus on code review generation, aiming to further explore ways to improve LLM performance on this task.

\subsection{Large Language Model}
\label{subsection:largeLanguageModel}

Language models have evolved from early statistical models to neural, pre-trained models, and now to large language models (LLMs) with over 10B parameters, trained on vast datasets \cite{zhao2023survey}. LLMs show superior performance and emergent capabilities \cite{wei2022emergent}, often matching or surpassing task-specific pre-trained models. Notable LLMs include ChatGPT-4o, which excels across benchmarks \cite{openAIGPT4o}, and DeepSeek v3 \cite{liu2024deepseek} among open-source models. Our research builds on ChatGPT-4o and DeepSeek v3, applying the LAURA framework to both.

So far, LLMs have been widely used in software engineering tasks like vulnerability prediction \cite{sun2024gptscan, zhou2024large, sagodi2024reality} and commit message generation \cite{eliseeva2023commit, wu2024commit, li2024only}. Many studies have explored ways to improve the performance of LLMs on specific tasks. Generally, LLMs can be further enhanced through methods such as augmenting input information \cite{lu2024exploring}, retrieval-augmented generation \cite{gao2023retrieval}, and prompt engineering \cite{white2024chatgpt}. Augmenting input information is based on the intuitive idea that providing the model with richer contextual information helps it better understand the task and produce better results. Retrieval-augmented generation \cite{lewis2020retrieval} enhances the model's task awareness by incorporating knowledge from external databases, thereby strengthening the model's generation capabilities. Prompt engineering \cite{liu2023pre} involves designing prompts for the model, guiding it through instructions on the actions it must perform with the given input, and ensuring that the LLMs achieve high performance in downstream tasks.


\section{Conclusion}
\label{section:conclusion}

In this paper, we introduce LAURA, a novel LLM-based generation framework for code review, which enhances the quality of code review comments generated by LLMs through three components: context augmentation, review exemplar retrieval, and systematic guidance. We applied LAURA to ChatGPT-4o and DeepSeek v3, and demonstrated through both LLM-based and human evaluations that LAURA significantly improves their performance on code review generation tasks. Experimental results show that LAURA outperforms direct use of LLMs for code review generation and substantially surpasses the performance of the pretrained model CodeReviewer.

Furthermore, our findings indicate that each of the three components -- context augmentation, review exemplar retrieval, and systematic guidance -- contributes to improving LLM performance in code review generation tasks, with the combination of all three yielding the best results. To further enhance the quality of generated code review comments, future research should explore additional useful input contexts and better methods for input integration.

\section*{Acknowledgments}
This work is supported by the National Natural Science Foundation of China Grants (62202048, 62232003, 62141209, 62402482, and 62572048).

\balance
\bibliographystyle{ieeetr}
\bibliography{references}

@String{Computing = "Computing" }

@article{golzadeh2021ground,
title       ={A ground-truth dataset and classification model for detecting bots in GitHub issue and PR comments},
author      ={Golzadeh, Mehdi and Decan, Alexandre and Legay, Damien and Mens, Tom},
journal     ={Journal of Systems and Software},
volume      ={175},
pages       ={110911},
year        ={2021},
publisher   ={Elsevier}
}

@inproceedings{mahajan2020recommending,
title       ={Recommending stack overflow posts for fixing runtime exceptions using failure scenario matching},
author      ={Mahajan, Sonal and Abolhassani, Negarsadat and Prasad, Mukul R},
booktitle   ={Proceedings of the 28th ACM Joint Meeting on European Software Engineering Conference and Symposium on the Foundations of Software Engineering},
pages       ={1052--1064},
year        ={2020}
}

@inproceedings{tufano2022using,
title       ={Using pre-trained models to boost code review automation},
author      ={Tufano, Rosalia and Masiero, Simone and Mastropaolo, Antonio and Pascarella, Luca and Poshyvanyk, Denys and Bavota, Gabriele},
booktitle   ={Proceedings of the 44th international conference on software engineering},
pages       ={2291--2302},
year        ={2022}
}

@inproceedings{sadowski2018modern,
  title={Modern code review: a case study at google},
  author={Sadowski, Caitlin and S{\"o}derberg, Emma and Church, Luke and Sipko, Michal and Bacchelli, Alberto},
  booktitle={Proceedings of the 40th international conference on software engineering: Software engineering in practice},
  pages={181--190},
  year={2018}
}

@inproceedings{zhou2017scalability,
  title={On the scalability of linux kernel maintainers' work},
  author={Zhou, Minghui and Chen, Qingying and Mockus, Audris and Wu, Fengguang},
  booktitle={Proceedings of the 2017 11th joint meeting on foundations of software engineering},
  pages={27--37},
  year={2017}
}

@inproceedings{mcintosh2014impact,
  title={The impact of code review coverage and code review participation on software quality: A case study of the qt, vtk, and itk projects},
  author={McIntosh, Shane and Kamei, Yasutaka and Adams, Bram and Hassan, Ahmed E},
  booktitle={Proceedings of the 11th working conference on mining software repositories},
  pages={192--201},
  year={2014}
}

@inproceedings{bavota2015four,
  title={Four eyes are better than two: On the impact of code reviews on software quality},
  author={Bavota, Gabriele and Russo, Barbara},
  booktitle={2015 IEEE International Conference on Software Maintenance and Evolution (ICSME)},
  pages={81--90},
  year={2015},
  organization={IEEE}
}

@article{wang2023codet5plus,
title       ={Codet5+: Open code large language models for code understanding and generation},
author      ={Wang, Yue and Le, Hung and Gotmare, Akhilesh Deepak and Bui, Nghi DQ and Li, Junnan and Hoi, Steven CH},
journal     ={arXiv preprint arXiv:2305.07922},
year        ={2023}
}

@article{kartal2024automating,
title       ={Automating modern code review processes with code similarity measurement},
author      ={Kartal, Yusuf and Akdeniz, E Kaan and {\"O}zkan, Kemal},
journal     ={Information and Software Technology},
volume      ={173},
pages       ={107490},
year        ={2024},
publisher   ={Elsevier}
}

@inproceedings{wang2021codet5,
title       ={CodeT5: Identifier-aware Unified Pre-trained Encoder-Decoder Models for Code Understanding and Generation}, 
author      ={Wang, Yue and Wang, Weishi and Joty, Shafiq and Hoi, Steven CH},
booktitle   ={EMNLP},
year        ={2021}
}

@article{tufano2024code,
title       ={Code review automation: strengths and weaknesses of the state of the art},
author      ={Tufano, Rosalia and Dabi{\'c}, Ozren and Mastropaolo, Antonio and Ciniselli, Matteo and Bavota, Gabriele},
journal     ={IEEE Transactions on Software Engineering},
year        ={2024},
publisher   ={IEEE}
}

@inproceedings{thongtanunam2015should,
title       ={Who should review my code? a file location-based code-reviewer recommendation approach for modern code review},
author      ={Thongtanunam, Patanamon and Tantithamthavorn, Chakkrit and Kula, Raula Gaikovina and Yoshida, Norihiro and Iida, Hajimu and Matsumoto, Ken-ichi},
booktitle   ={2015 IEEE 22nd International Conference on Software Analysis, Evolution, and Reengineering (SANER)},
pages       ={141--150},
year        ={2015},
organization={IEEE}
}

@article{chouchen2021whoreview,
title       ={WhoReview: A multi-objective search-based approach for code reviewers recommendation in modern code review},
author      ={Chouchen, Moataz and Ouni, Ali and Mkaouer, Mohamed Wiem and Kula, Raula Gaikovina and Inoue, Katsuro},
journal     ={Applied Soft Computing},
volume      ={100},
pages       ={106908},
year        ={2021},
publisher   ={Elsevier}
}

@inproceedings{hong2022commentfinder,
title       ={Commentfinder: a simpler, faster, more accurate code review comments recommendation},
author      ={Hong, Yang and Tantithamthavorn, Chakkrit and Thongtanunam, Patanamon and Aleti, Aldeida},
booktitle   ={Proceedings of the 30th ACM joint European software engineering conference and symposium on the foundations of software engineering},
pages       ={507--519},
year        ={2022}
}

@inproceedings{shuvo2023recommending,
title       ={Recommending code reviews leveraging code changes with structured information retrieval},
author      ={Shuvo, Ohiduzzaman and Mahbub, Parvez and Rahman, Mohammad Masudur},
booktitle   ={2023 IEEE International Conference on Software Maintenance and Evolution (ICSME)},
pages       ={194--206},
year        ={2023},
organization={IEEE}
}

@inproceedings{li2022automating,
title       ={Automating code review activities by large-scale pre-training},
author      ={Li, Zhiyu and Lu, Shuai and Guo, Daya and Duan, Nan and Jannu, Shailesh and Jenks, Grant and Majumder, Deep and Green, Jared and Svyatkovskiy, Alexey and Fu, Shengyu and others},
booktitle   ={Proceedings of the 30th ACM Joint European Software Engineering Conference and Symposium on the Foundations of Software Engineering},
pages       ={1035--1047},
year        ={2022}
}

@inproceedings{lu2023llama,
title       ={LLaMA-Reviewer: Advancing Code Review Automation with Large Language Models through Parameter-Efficient Fine-Tuning},
author      ={Lu, Junyi and Yu, Lei and Li, Xiaojia and Yang, Li and Zuo, Chun},
booktitle   ={2023 IEEE 34th International Symposium on Software Reliability Engineering (ISSRE)},
pages       ={647--658},
year        ={2023},
organization={IEEE}
}

@inproceedings{hellendoorn2021towards,
title       ={Towards automating code review at scale},
author      ={Hellendoorn, Vincent J and Tsay, Jason and Mukherjee, Manisha and Hirzel, Martin},
booktitle   ={Proceedings of the 29th ACM Joint Meeting on European Software Engineering Conference and Symposium on the Foundations of Software Engineering},
pages       ={1479--1482},
year        ={2021}
}

@inproceedings{tufano2021towards,
title       ={Towards automating code review activities},
author      ={Tufano, Rosalia and Pascarella, Luca and Tufano, Michele and Poshyvanyk, Denys and Bavota, Gabriele},
booktitle   ={2021 IEEE/ACM 43rd International Conference on Software Engineering (ICSE)},
pages       ={163--174},
year        ={2021},
organization={IEEE}
}

@inproceedings{kononenko2016code,
title       ={Code review quality: How developers see it},
author      ={Kononenko, Oleksii and Baysal, Olga and Godfrey, Michael W},
booktitle   ={Proceedings of the 38th international conference on software engineering},
pages       ={1028--1038},
year        ={2016}
}

@article{turzo2024makes,
title       ={What makes a code review useful to opendev developers? an empirical investigation},
author      ={Turzo, Asif Kamal and Bosu, Amiangshu},
journal     ={Empirical Software Engineering},
volume      ={29},
number      ={1},
pages       ={6},
year        ={2024},
publisher   ={Springer}
}

@inproceedings{sulun2019suggesting,
title       ={Suggesting reviewers of software artifacts using traceability graphs},
author      ={S{\"u}l{\"u}n, Emre},
booktitle   ={Proceedings of the 2019 27th ACM Joint Meeting on European Software Engineering Conference and Symposium on the Foundations of Software Engineering},
pages       ={1250--1252},
year        ={2019}
}

@inproceedings{asthana2019whodo,
title       ={Whodo: Automating reviewer suggestions at scale},
author      ={Asthana, Sumit and Kumar, Rahul and Bhagwan, Ranjita and Bird, Christian and Bansal, Chetan and Maddila, Chandra and Mehta, Sonu and Ashok, Balasubramanyan},
booktitle   ={Proceedings of the 2019 27th ACM joint meeting on european software engineering conference and symposium on the foundations of software engineering},
pages       ={937--945},
year        ={2019}
}

@inproceedings{pandya2022corms,
title={Corms: a github and gerrit based hybrid code reviewer recommendation approach for modern code review},
author={Pandya, Prahar and Tiwari, Saurabh},
booktitle={Proceedings of the 30th ACM joint European software engineering conference and symposium on the foundations of software engineering},
pages={546--557},
year={2022}
}

@inproceedings{kong2022recommending,
title       ={Recommending code reviewers for proprietary software projects: A large scale study},
author      ={Kong, Dezhen and Chen, Qiuyuan and Bao, Lingfeng and Sun, Chenxing and Xia, Xin and Li, Shanping},
booktitle   ={2022 IEEE International Conference on Software Analysis, Evolution and Reengineering (SANER)},
pages       ={630--640},
year        ={2022},
organization={IEEE}
}

@inproceedings{mirsaeedi2020mitigating,
title       ={Mitigating turnover with code review recommendation: Balancing expertise, workload, and knowledge distribution},
author      ={Mirsaeedi, Ehsan and Rigby, Peter C},
booktitle   ={Proceedings of the ACM/IEEE 42nd international conference on software engineering},
pages       ={1183--1195},
year        ={2020}
}

@inproceedings{chueshev2020expanding,
title       ={Expanding the number of reviewers in open-source projects by recommending appropriate developers},
author      ={Chueshev, Aleksandr and Lawall, Julia and Bendraou, Reda and Ziadi, Tewfik},
booktitle   ={2020 IEEE International Conference on Software Maintenance and Evolution (ICSME)},
pages       ={499--510},
year        ={2020},
organization={IEEE}
}

@article{raffel2020exploring,
title       ={Exploring the limits of transfer learning with a unified text-to-text transformer},
author      ={Raffel, Colin and Shazeer, Noam and Roberts, Adam and Lee, Katherine and Narang, Sharan and Matena, Michael and Zhou, Yanqi and Li, Wei and Liu, Peter J},
journal     ={Journal of machine learning research},
volume      ={21},
number      ={140},
pages       ={1--67},
year        ={2020}
}

@article{wei2022emergent,
title       ={Emergent abilities of large language models},
author      ={Wei, Jason and Tay, Yi and Bommasani, Rishi and Raffel, Colin and Zoph, Barret and Borgeaud, Sebastian and Yogatama, Dani and Bosma, Maarten and Zhou, Denny and Metzler, Donald and others},
journal     ={arXiv preprint arXiv:2206.07682},
year        ={2022}
}

@article{zhao2023survey,
title       ={A survey of large language models},
author      ={Zhao, Wayne Xin and Zhou, Kun and Li, Junyi and Tang, Tianyi and Wang, Xiaolei and Hou, Yupeng and Min, Yingqian and Zhang, Beichen and Zhang, Junjie and Dong, Zican and others},
journal     ={arXiv preprint arXiv:2303.18223},
year        ={2023}
}

@inproceedings{eliseeva2023commit,
title       ={From commit message generation to history-aware commit message completion},
author      ={Eliseeva, Aleksandra and Sokolov, Yaroslav and Bogomolov, Egor and Golubev, Yaroslav and Dig, Danny and Bryksin, Timofey},
booktitle   ={2023 38th IEEE/ACM International Conference on Automated Software Engineering (ASE)},
pages       ={723--735},
year        ={2023},
organization={IEEE}
}

@inproceedings{wu2024commit,
title       ={Commit Message Generation via ChatGPT: How Far Are We?},
author      ={Wu, Yifan and Li, Ying and Yu, Siyu},
booktitle   ={Proceedings of the 2024 IEEE/ACM First International Conference on AI Foundation Models and Software Engineering},
pages       ={124--129},
year        ={2024}
}

@article{li2024only,
title       ={Only diff is not enough: Generating commit messages leveraging reasoning and action of large language model},
author      ={Li, Jiawei and Farag{\'o}, David and Petrov, Christian and Ahmed, Iftekhar},
journal     ={Proceedings of the ACM on Software Engineering},
volume      ={1},
number      ={FSE},
pages       ={745--766},
year        ={2024},
publisher   ={ACM New York, NY, USA}
}

@inproceedings{sun2024gptscan,
title       ={Gptscan: Detecting logic vulnerabilities in smart contracts by combining gpt with program analysis},
author      ={Sun, Yuqiang and Wu, Daoyuan and Xue, Yue and Liu, Han and Wang, Haijun and Xu, Zhengzi and Xie, Xiaofei and Liu, Yang},
booktitle   ={Proceedings of the IEEE/ACM 46th International Conference on Software Engineering},
pages       ={1--13},
year        ={2024}
}

@inproceedings{zhou2024large,
title       ={Large language model for vulnerability detection: Emerging results and future directions},
author      ={Zhou, Xin and Zhang, Ting and Lo, David},
booktitle   ={Proceedings of the 2024 ACM/IEEE 44th International Conference on Software Engineering: New Ideas and Emerging Results},
pages       ={47--51},
year        ={2024}
}

@inproceedings{sagodi2024reality,
title       ={Reality Check: Assessing GPT-4 in Fixing Real-World Software Vulnerabilities},
author      ={S{\'a}godi, Zolt{\'a}n and Antal, G{\'a}bor and Bogenf{\"u}rst, Bence and Isztin, Martin and Heged{\H{u}}s, P{\'e}ter and Ferenc, Rudolf},
booktitle   ={Proceedings of the 28th International Conference on Evaluation and Assessment in Software Engineering},
pages       ={252--261},
year        ={2024}
}

@inproceedings{guo2024exploring,
title       ={Exploring the potential of chatgpt in automated code refinement: An empirical study},
author      ={Guo, Qi and Cao, Junming and Xie, Xiaofei and Liu, Shangqing and Li, Xiaohong and Chen, Bihuan and Peng, Xin},
booktitle   ={Proceedings of the 46th IEEE/ACM International Conference on Software Engineering},
pages       ={1--13},
year        ={2024}
}

@article{pornprasit2024fine,
title       ={Fine-tuning and prompt engineering for large language models-based code review automation},
author      ={Pornprasit, Chanathip and Tantithamthavorn, Chakkrit},
journal     ={Information and Software Technology},
volume      ={175},
pages       ={107523},
year        ={2024},
publisher   ={Elsevier}
}

@article{lu2024exploring,
title       ={Exploring the impact of code review factors on the code review comment generation},
author      ={Lu, Junyi and Li, Zhangyi and Shen, Chenjie and Yang, Li and Zuo, Chun},
journal     ={Automated Software Engineering},
volume      ={31},
number      ={2},
pages       ={71},
year        ={2024},
publisher   ={Springer}
}

@article{gao2023retrieval,
title       ={Retrieval-augmented generation for large language models: A survey},
author      ={Gao, Yunfan and Xiong, Yun and Gao, Xinyu and Jia, Kangxiang and Pan, Jinliu and Bi, Yuxi and Dai, Yi and Sun, Jiawei and Wang, Meng and Wang, Haofen},
journal     ={arXiv preprint arXiv:2312.10997},
year        ={2023}
}

@incollection{white2024chatgpt,
title       ={Chatgpt prompt patterns for improving code quality, refactoring, requirements elicitation, and software design},
author      ={White, Jules and Hays, Sam and Fu, Quchen and Spencer-Smith, Jesse and Schmidt, Douglas C},
booktitle   ={Generative AI for Effective Software Development},
pages       ={71--108},
year        ={2024},
publisher   ={Springer}
}

@article{lewis2020retrieval,
title       ={Retrieval-augmented generation for knowledge-intensive nlp tasks},
author      ={Lewis, Patrick and Perez, Ethan and Piktus, Aleksandra and Petroni, Fabio and Karpukhin, Vladimir and Goyal, Naman and K{\"u}ttler, Heinrich and Lewis, Mike and Yih, Wen-tau and Rockt{\"a}schel, Tim and others},
journal     ={Advances in Neural Information Processing Systems},
volume      ={33},
pages       ={9459--9474},
year        ={2020}
}

@article{liu2023pre,
title       ={Pre-train, prompt, and predict: A systematic survey of prompting methods in natural language processing},
author      ={Liu, Pengfei and Yuan, Weizhe and Fu, Jinlan and Jiang, Zhengbao and Hayashi, Hiroaki and Neubig, Graham},
journal     ={ACM Computing Surveys},
volume      ={55},
number      ={9},
pages       ={1--35},
year        ={2023},
publisher   ={ACM New York, NY}
}

@inproceedings{tian2022what,
author      ={Tian, Yingchen and Zhang, Yuxia and Stol, Klaas-Jan and Jiang, Lin and Liu, Hui},
title       ={What makes a good commit message?},
year        ={2022},
isbn        ={9781450392211},
publisher   ={Association for Computing Machinery},
address     ={New York, NY, USA},
url         ={https://doi.org/10.1145/3510003.3510205},
doi         ={10.1145/3510003.3510205},
booktitle   ={Proceedings of the 44th International Conference on Software Engineering},
pages       ={2389–2401},
numpages    ={13},
location    ={Pittsburgh, Pennsylvania},
series      ={ICSE '22}
}

@article{zhang2024rag,
title       ={RAG-Enhanced Commit Message Generation},
author      ={Zhang, Linghao and Zhang, Hongyi and Wang, Chong and Liang, Peng},
journal     ={arXiv preprint arXiv:2406.05514},
year        ={2024}
}

@article{wang2021context,
title       ={Context-aware retrieval-based deep commit message generation},
author      ={Wang, Haoye and Xia, Xin and Lo, David and He, Qiang and Wang, Xinyu and Grundy, John},
journal     ={ACM Transactions on Software Engineering and Methodology (TOSEM)},
volume      ={30},
number      ={4},
pages       ={1--30},
year        ={2021},
publisher   ={ACM New York, NY, USA}
}

@article{bae2024enhancing,
title       ={Enhancing Software Code Vulnerability Detection Using GPT-4o and Claude-3.5 Sonnet: A Study on Prompt Engineering Techniques},
author      ={Bae, Jaehyeon and Kwon, Seoryeong and Myeong, Seunghwan},
journal     ={Electronics},
volume      ={13},
number      ={13},
pages       ={2657},
year        ={2024},
publisher   ={MDPI}
}

@inproceedings{ahmed2024automatic,
title       ={Automatic semantic augmentation of language model prompts (for code summarization)},
author      ={Ahmed, Toufique and Pai, Kunal Suresh and Devanbu, Premkumar and Barr, Earl},
booktitle   ={Proceedings of the IEEE/ACM 46th International Conference on Software Engineering},
pages       ={1--13},
year        ={2024}
}

@article{white2023prompt,
title       ={A prompt pattern catalog to enhance prompt engineering with chatgpt},
author      ={White, Jules and Fu, Quchen and Hays, Sam and Sandborn, Michael and Olea, Carlos and Gilbert, Henry and Elnashar, Ashraf and Spencer-Smith, Jesse and Schmidt, Douglas C},
journal     ={arXiv preprint arXiv:2302.11382},
year        ={2023}
}

@article{amatriain2024prompt,
title       ={Prompt design and engineering: Introduction and advanced methods},
author      ={Amatriain, Xavier},
journal     ={arXiv preprint arXiv:2401.14423},
year        ={2024}
}

@article{fiske1992but,
title       ={But the reviewers are making different criticisms of my paper! Diversity and uniqueness in reviewer comments.},
author      ={Fiske, Donald W and Fogg, Louis},
year        ={1992},
publisher   ={American Psychological Association}
}

@article{reiter2018structured,
title       ={A structured review of the validity of BLEU},
author      ={Reiter, Ehud},
journal     ={Computational Linguistics},
volume      ={44},
number      ={3},
pages       ={393--401},
year        ={2018},
publisher   ={MIT Press One Rogers Street, Cambridge, MA 02142-1209, USA journals-info~…}
}

@inproceedings{dyer2013boa,
title       ={Boa: A language and infrastructure for analyzing ultra-large-scale software repositories},
author      ={Dyer, Robert and Nguyen, Hoan Anh and Rajan, Hridesh and Nguyen, Tien N},
booktitle   ={2013 35th International Conference on Software Engineering (ICSE)},
pages       ={422--431},
year        ={2013},
organization={IEEE}
}

@inproceedings{schall2024commitbench,
title       ={CommitBench: A benchmark for commit message generation},
author      ={Schall, Maxmilian and Czinczoll, Tamara and De Melo, Gerard},
booktitle   ={2024 IEEE International Conference on Software Analysis, Evolution and Reengineering (SANER)},
pages       ={728--739},
year        ={2024},
organization={IEEE}
}

@article{ochiai1957zoogeographic,
title       ={Zoogeographic studies on the soleoid fishes found in Japan and its neighbouring regions},
author      ={Ochiai, Akira},
journal     ={Bulletin of Japanese Society of Scientific Fisheries},
volume      ={22},
pages       ={526--530},
year        ={1957}
}

@article{liu2024deepseek,
title={Deepseek-v3 technical report},
author={Liu, Aixin and Feng, Bei and Xue, Bing and Wang, Bingxuan and Wu, Bochao and Lu, Chengda and Zhao, Chenggang and Deng, Chengqi and Zhang, Chenyu and Ruan, Chong and others},
journal={arXiv preprint arXiv:2412.19437},
year={2024}
}

@inproceedings{bosu2015characteristics,
title={Characteristics of useful code reviews: An empirical study at microsoft},
author={Bosu, Amiangshu and Greiler, Michaela and Bird, Christian},
booktitle={2015 IEEE/ACM 12th Working Conference on Mining Software Repositories},
pages={146--156},
year={2015},
organization={IEEE}
}

@article{rong2024distilling,
title={Distilling Quality Enhancing Comments From Code Reviews to Underpin Reviewer Recommendation},
author={Rong, Guoping and Yu, Yongda and Zhang, Yifan and Zhang, He and Shen, Haifeng and Shao, Dong and Kuang, Hongyu and Wang, Min and Wei, Zhao and Xu, Yong and others},
journal={IEEE Transactions on Software Engineering},
volume={50},
number={7},
pages={1658--1674},
year={2024},
publisher={IEEE}
}

@inproceedings{lu2025deepcrceval,
title={DeepCRCEval: Revisiting the Evaluation of Code Review Comment Generation},
author={Lu, Junyi and Li, Xiaojia and Hua, Zihan and Yu, Lei and Cheng, Shiqi and Yang, Li and Zhang, Fengjun and Zuo, Chun},
booktitle={International Conference on Fundamental Approaches to Software Engineering},
pages={43--64},
year={2025},
organization={Springer}
}

@misc{LAURAfigshare,
author      ={Zhang, Yuxin and Zhang, Yuxia and Sun, Zeyu and Jiang, Yanjie and Liu, Hui},
title       ={LAURA: Enhancing Code Review Generation with Context-Enriched Retrieval-Augmented LLM},
howpublished={Figshare},
year        ={2025},
note        ={\url{https://doi.org/10.6084/m9.figshare.27367194}}
}

@misc{openAIGPT4o,
author      ={OpenAI},
title       ={Hello GPT-4o | OpenAI},
howpublished={Website},
year        ={2024},
note        ={\url{https://openai.com/index/hello-gpt-4o/}}
}

@misc{JonathanCorbet,
author      ={Jonathan Corbet},
title       ={On Linux kernel maintainer scalability [LWN.net]},
howpublished={Website},
year        ={2016},
note        ={\url{https://lwn.net/Articles/703005/}}
}

@misc{Tree-sitter,
author      ={Tree-sitter},
title       ={Introduction - Tree-sitter},
howpublished={Website},
year        ={2025},
note        ={\url{https://tree-sitter.github.io/tree-sitter/}}
}

@misc{GitHubGraphQL,
author      ={GitHub},
title       ={GitHub GraphQL API documentation - GitHub Docs},
howpublished={Website},
year        ={2025},
note        ={\url{https://docs.github.com/en/graphql}}
}

@misc{GitHubCompare,
author      ={GitHub},
title       ={Comparing commits - GitHub Docs},
howpublished={Website},
year        ={2025},
note        ={\url{https://docs.github.com/en/pull-requests/committing-changes-to-your-project/viewing-and-comparing-commits/comparing-commits}}
}

@article{cohen1960coefficient,
title={A coefficient of agreement for nominal scales},
author={Cohen, Jacob},
journal={Educational and psychological measurement},
volume={20},
number={1},
pages={37--46},
year={1960},
publisher={Sage Publications Sage CA: Thousand Oaks, CA}
}

@article{joshi2015likert,
title={Likert scale: Explored and explained},
author={Joshi, Ankur and Kale, Saket and Chandel, Satish and Pal, D Kumar},
journal={British journal of applied science \& technology},
volume={7},
number={4},
pages={396},
year={2015},
publisher={Sciencedomain International}
}

\end{document}